# Examining quality of DGNSS derived positioning in data in urban city- A case study of an urban city in India

*Jai G. Singla, Sunanda Trivedi*

[1]*Space Applications Centre, ISRO, Ahmedabad – 380015*
*Corresponding author. e-mail: jaisingla@gmail.com

Abstract:

GNSS observations are carried out in static mode/ Differential global navigation satellite system (DGNSS) and dynamic mode / Real time Kinematics (RTK) mainly. RTK mode of observation is useful in case of navigation whereas in order to determine very precise positioning, static / DGNSS/ DGPS mode is recommended. In this study, we have examined the quality of DGNSS survey of an urban city in India over ~300 Ground Control Points. Survey is carried out in DGNSS mode with dual frequency mode. All the observations were recorded using GPS, GLONASS , Galileo and Beidu with GDOP values in the range of 1.4 to 2.5. Beidu was used in broadcast ephemeris mode whereas for other constellations, precise orbit ephemeris were obtained from International GNSS service (IGS) site as per the observation day and month. Further, all the data was post processed in the SW suite and positional and vertical accuracies of millimeter to few centimeter level were obtained. This paper describes the approach of Ground Control Point (GCP) identification, surveying, methodology, use of CORS network and data post-processing in order to achieve such a precise accuracies in the urban city.

## 1. Introduction

GNSS is widely used in various applications such as surveying and mapping , vehicle tracking and management , precision agriculture, GNSS Meteorology, GNSS Ionosphere and space weather, GNSS reflectometry, earth quake monitoring, land and structural health monitoring etc. (Dardanelli et al, 2022). GPS observations are made in several ways i.e. Precise Point Positioning (PPP) using a single frequency based receiver whereas relative positioning involves two GPS receivers (Rao , 2010) . PPP is mainly used for navigation purpose and relative mode is used for high precision applications. Further, static surveying and real time kinematic (RTK) surveying falls under relative mode of observation. In static / DGNSS / DGPS mode, reference and rover modules are kept stationary. Reference station is equipped with a stationary receiver which is continuously receiving the satellite signals and calculates the position of the reference. Whereas, in order to obtain precise location of the rover receivers, static observations minimum for 45 minutes are recorded at rover end in order to cancel User Equivalent Range Errors (UERE) . In case of RTK, reference receiver is fixed at one location and rover is continuously moving from one place to another.

As receiver on the ground is recording observations, there are many sources of errors such as clock, orbit ephemeris, ionosphere, troposphere, and multipath. **User Equivalent Range Error (UERE)** is square root of sum of squares of individual biases from Ephemeris, clock, Ionosphere, Troposphere and multipath. All these biases can be minimized by better DOP values and longer observation times using dual frequency observations. Currently, RTK positioning is being used as one of the most popular techniques for real-time precise positioning using Global Navigation Satellite Systems (GNSS) carrier phase observations. Although RTK can be used in many fields like surveying and navigation. But, it is challenging to obtain precise positioning using RTK in dense urban areas. Due to tall and dense buildings in the urban area satellite signals are blocked, hence creates multi-path error. Therefore, in order to obtain precise centimeter to millimeter level accuracies in urban areas, Differential GNSS / DGPS / static surveying is recommended.

In India, Survey of India (SOI) has established the Continuously Operating Reference Stations System (CORS) with an accuracy of + / - 3cm (CORS site). CORS network is the fundamental infrastructure to correct the rover observations all over the country. In our study, Gandhinagar reference station has been used to correct many observations.

**Literature survey**

From last two- three decades, many researchers across the globe worked on Global Navigation Satellite System, its configurations, algorithms, accuracies in various modes. Richard Langley article on Dilution of Precision (DOP) provides insights about basic understanding of concepts related to Global Positioning System (GPS) terminologies. Alkan et al, 2015 performed accuracy assessment of Precise Point Positioning (PPP) technique using GPS and GLONASS in urban area. Difference in position vary between 2-18 cm for GPS only, and 3-8cm for GPS and GLONASS combined. Dabove et al , 2018 assessed the performance of GPS-GLONASS and GPS-BDS in Europe. They achieved centimetre level accuracies using both the constellations for static positioning.

A few researchers worked on accuracies of GNSS in urban canopy. Xu et al, 2018 carried out the performance analysis in urban areas using GPS/ BDS dual / triple frequency NRTK and discussed the issues that affect the performance in urban areas. They concluded that BDS triple frequency observation significantly improved the initialization time and positional accuracy of RTK in Hong Kong. Ma et al, 2020 studied the urban road GPS environment with respect to bus routes and trajectories in urban city. They obtained GPS environment friendliness (GEF) results from poor to satisfactory range.

In Indian scenario, Goswami et al 2021 studied the GPS and GPS-GLONASS combined accuracies for Indian defence usage. They concluded that GPS+GLONASS combination will give better position solution in case missile maintains a constant altitude. Sundara et al, 2022 estimated accuracy of Indian Regional Navigational Satellite System (IRNSS) in different configurations. They obtained 2DRMS values of 3- 4m using differential NAVIC in two different configurations. Leena et al , 2023 assessed positional accuracies of LIDAR based point cloud with respect to GCPs. They observed more shift in vertical direction (height) as compared to horizontal direction. Anurag et al 2023, assessed the accuracy of relative GPS as function of distance and duration for CORS network. The concluded that up to 160KM, there is a negligible dependency of the accuracy over distance.

Our study is first of its kind of study conducted in India as per best of our knowledge. In this study, static dual frequency observations were recorded over ~ 300 points on a dense city area containing all types (small/ mid-size and high rise) of buildings. These static observations were made using GPS, GLONASS , Galileo and Beidu with GDOP values mostly in the range of 1.4 to 2.5. Beidu was used in broadcast ephemeris mode. Whereas for other constellations, precise orbit ephemeris were obtained from International GNSS service (IGS) site as per the observation day and month. Further, all the data was post processed using three different approaches as per Section-3. All the points were processed, examined and positional & vertical accuracies ranging from millimeter to few centimeter levels were obtained.

## 2. Study Area

Our study consists of an area of 25 square KM area of Gandhinagar city and surroundings. Gandhinagar is the capital of Gujarat state and contains 326 $〚km〛^2$ area. Gandhinagar is chosen for the study due to its density and availability of all type of building structures. Apart from that, Gandhinagar also contain CORS reference point. The extent of the area selected for surveying are : xMin, yMin 72.5967,23.1519 : xMax, yMax 72.6956,23.2728.

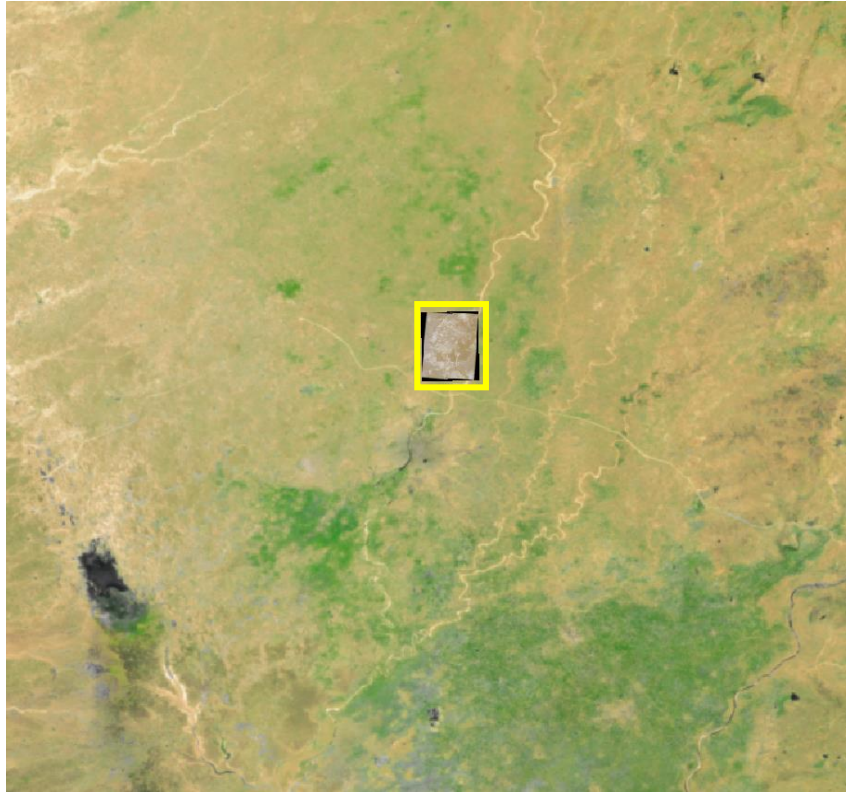

Figure-1 Study Area as shown in rectangular boundary.

## 3. Methodology

Methodology involves the steps of GCP identification and verification, Field survey planning and recording of DGPS observations as per the availability of Reference / rover stations and GCP post processing in order to achieve precise results.

### 3.1 GCP Identification

In a region of 10 KM in Length and 7 KM wide, well distributed GCPs are identified using the latest satellite datasets covering our area of interest (AOI). To ensure required distribution and proper point positioning, Bhuvan & Google images are also used. GCP distribution is done in such a way that it covers entire Gandhinagar area properly. At some location few close points are selected to have sufficient distribution for model points and check points. All identified points are marked in the image as shown in Figure-2. For verification of all the GCPs, photograph of GCP ground position and photograph with GPS receiver are obtained for records (Figure-3).

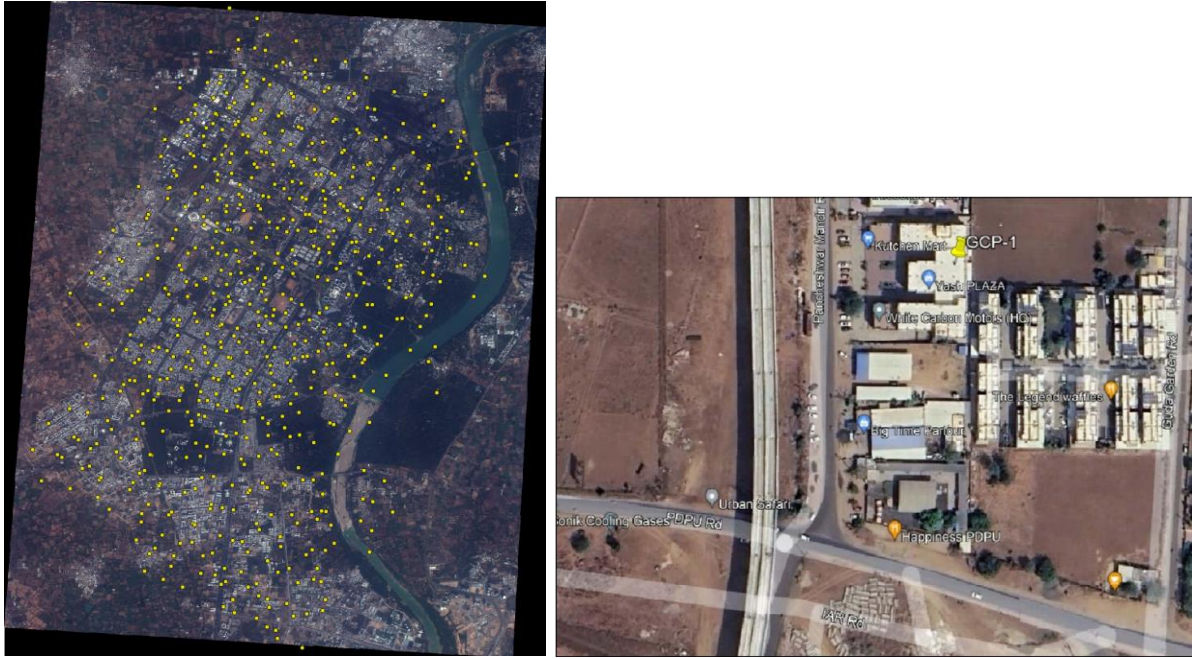

Figure-2 Identification of GCP points using latest satellite data

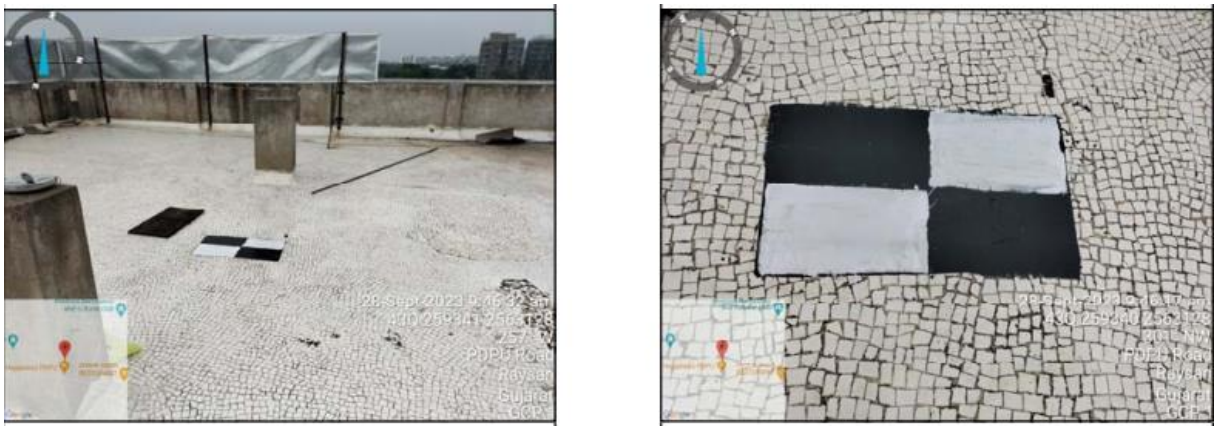

Figure 3 (Ground photographs of one of the GCP point)

**3.2 DGPS Field Survey Planning and approach**

DGPS surveying requires availability of reference as well as rover stations to carry out observations. Before approaching for the field survey, proper planning was carried out for GCP measurements. CORS point located at Gandhinagar was made as reference / zeroth point for our setup. Further, entire Gandhinagar was divided in to four zones and 1$^{st}$ order reference points were defined i.e. one for each zone (Figure-4). Corresponding to each 1$^{st}$ order point, static surveying for minimum 45 minutes were carried for each of the rover point. Initially 12-14 hours measurements were carried in order to correct 1$^{st}$ order point's w.r.t. Zeroth order point and thereafter minimum 45 minutes observations were recorded for each of the point (2$^{nd}$ order points) around 1$^{st}$ order points. As per the GCP distribution, 10-15 second order points were recorded in a day. Beside city map, GCP marked satellite image prints are used for the field survey.

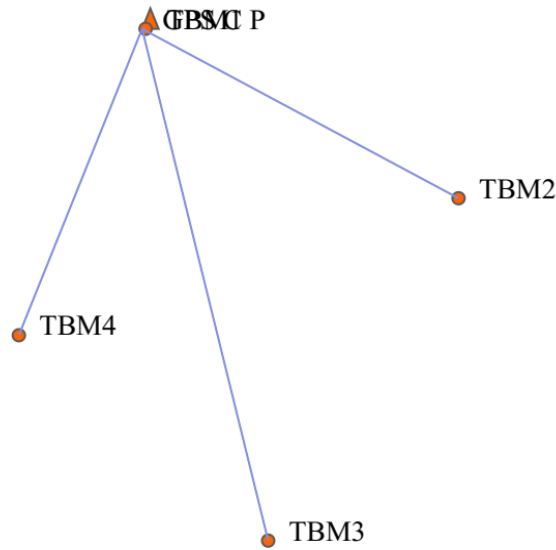

Figure-4 Establishment of 0<sup>th</sup> and 1<sup>st</sup> Order points and corresponding Baselines.

### 3.3 Differential mode Observations & Post Processing

In this study, GPS, GLONASS , Beidou and Galileo dual carrier phase observations are used to record data from all the satellites constellations.  DGPS / static mode of observation were used as explained below. Maximum linear distance of GCP reference point to Temporary Benchmark (TBM) was 6.5 KM.

**Differential GNSS/ GPS and Positional accuracy parameters**

In Differential GNSS technique, Reference station sends differential corrections to rover station. Differential corrections are calculated by current position and surveyed position of the reference positon. Afterwards rover station corrects its current position as per differential position.  (Sundara R.R. , 2022).

If reference system current positions is (X,Y,Z) and (X', Y', Z') are measured positions. Then differential position will be defined as subtraction of measure position from current position in all the three directions of X,Y and Z. Thereafter, differential positions are broadcasted to rover stations and rover's accurate position (rX,rY,rZ) is then computed using the current position (rX', rY' and rZ') and broadcasted differential corrections (dX, dY and dZ).

$dX = X-X'$ , $dY = Y-Y'$ , $dZ = Z-Z'$     - (1)

$rX = rX' + dX$, $rY = rY' + dY$ and $rZ = rZ' + dZ$     -(2)

Positional accuracy parameters:

Dilution of precision (DOP) is very important metric in order to define the accuracy and precision of the observation. Geometry dilution of the precision (GDOP) defines the position of the satellites w.r.t receiver. When more than four satellites are in the receiver's range, GDOP values are comparatively low and results into more accurate three dimensional position. Whereas **Horizontal dilution of the precision** (HDOP) is indicator of the accuracy on horizontal plane whereas **Positional dilution of precision** (PDOP) is defined as accuracy on horizontal and vertical place. In order to obtain better accuracies, Dual frequency receivers and maximum numbers of satellite visibility are essential.

$PDOP^2 = HDOP^2 + VDOP^2$ -(3)

$GDOP^2 = PDOP^2 + TDOP^2$ -(4)

Eq 4, TDOP stands for time DOP.

While carrying out the post processing exercise, we opted for three approaches.

- Data processing using identified base point.
- Data processing using CORS as base point.
- Online data processing using CORS base point.

**Data processing using identified base point.**

Initially, zeroth order reference point is processed first with 'SPP' single point processing mode with precise ephemeris data in limited edition of Leica infinity SW. Its ground coordinates are refined and updated with obtained coordinates. This point is now called as zeroth order point. Now, considering zeroth order point as reference point and other four Temporary Benchmark (TBM) points as rover points, baseline processing was performed with precise (millimetre level) accuracies (Figure-5). Here, precise accuracy is defined as the RMSE error in positional accuracy and vertical accuracies is obtained in millimetres. Thereafter, Each TBM point was made as reference point and each of the static point was made as rover point and data was processed to achieve precise coordinates and height accuracies for each of the point. By performing this exercise, we were able to process maximum number of points (~250 points) with millimetre level accuracies refer Figure-5.

**Data processing using Gandhinagar (GAAR) CORS as base point.**

Another mode of processing i.e. consideration of CORS reference station available at Gandhinagar as a reference point for processing was made to compute accuracies of remaining points. In this mode of processing, data from (CORS site) was obtained for the reference station. Further, processing of all the associated remaining points were carried out using the reference station data in Leica SW. Precise coordinates and height accuracies were obtained using this exercise.

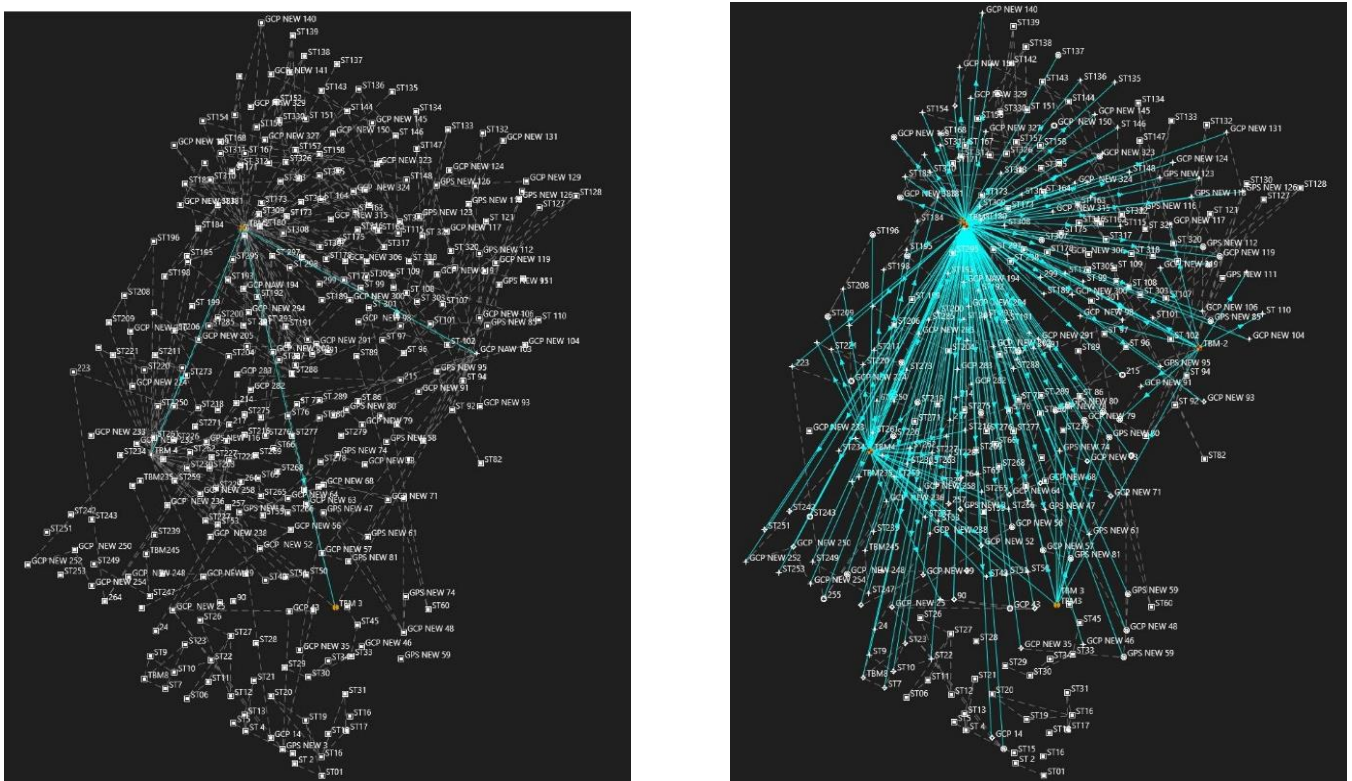

Figure-5 Processed Baseline using TBM points (Left), Processed Baselines for other rover points (Right)

**Online data processing using CORS base point.**

Due to some reason, CORS reference station data was not available for offline processing on a specific date. So, all the observations of that day were uploaded for processing on CORS website. CORS considers multiple (Nine) reference stations and calculated multiple baselines to obtain precise coordinates and height values for all the observations. All the remaining points were processed with millimetre to few centimetre level accuracies using this mode of processing too.

4.  **Results and Discussions**

    In the whole exercise, proper procedure of identification, observation and processing were followed. GDOP values of 1.3-1.5, PDOP 0.8-1.0, HDOP 0.4-0.5 and VDOP 0.7-0.8 values were recorded for maximum number of observations. Figure-6 depicts the DOP value and satellite visibility graph for one of the static points. From all the available/ visible satellite constellations during the time of observation, signals were obtained from 9/9 GPS

satellites, 18/23 Beidou satellites, 6/6 Galileo satellite (Figure-6). It was ensured to maintain a good dilution (1.3- 2.5) of precision for all the measurements.

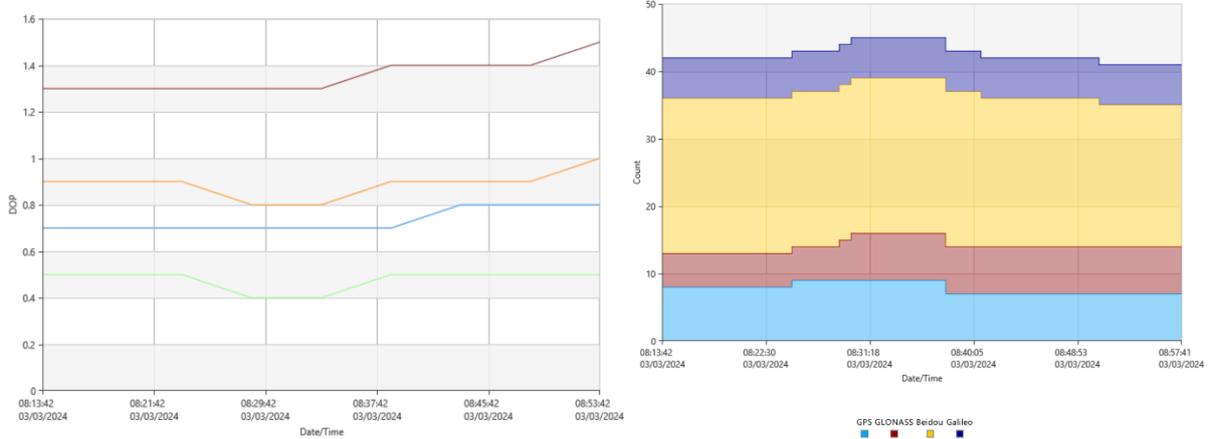

Figure -6 DOP Values and satellite visibility Graph for a point

After the GNSS processing as per the explained methodology, the height difference in navigated points vs fixed points (after corrections) were computed. It is seen from Figure-7 that correction in the height values varies from 0.5 meter to 12 meters range from navigated to fixed measurements. Apart from these differences, in Figure-8, error in third dimension (Height) is shown all over the corrected points. It is clearly evident from the plot that error over most of the point lies in millimetre whereas, for few of the points, it is computed up to few centimetres. Similarly, Figure-9 represents the final positional error value in one direction as well as in both the directions. Again, values of error lies in millimetre in most of the cases and reported in centimetre for few points.

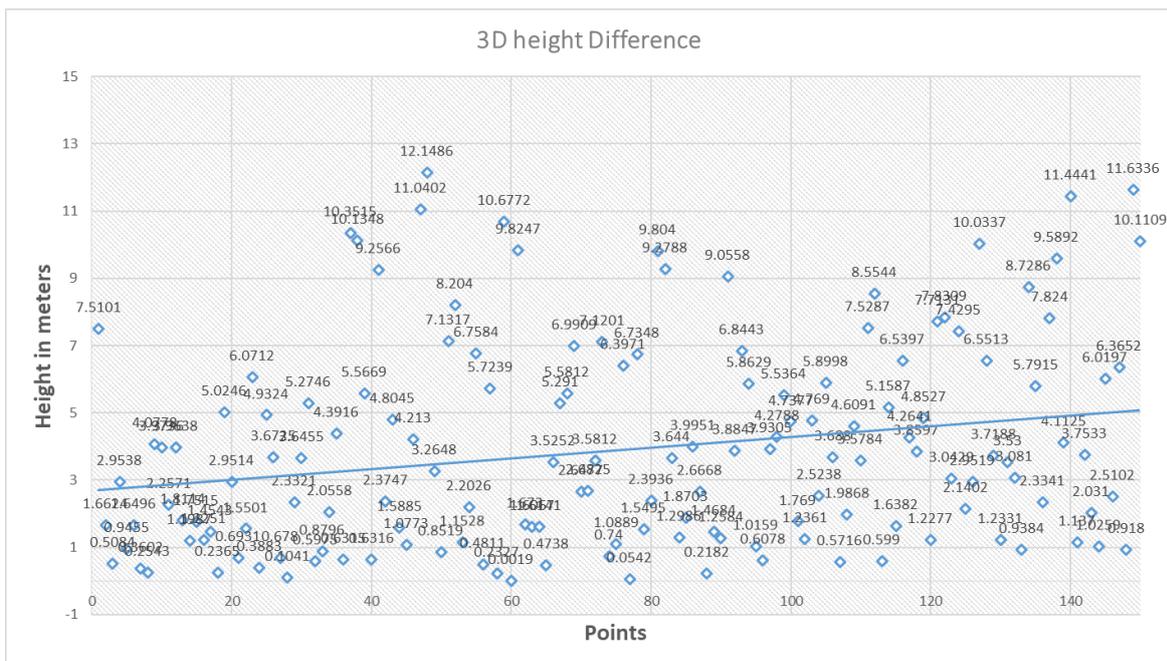

Figure-7 Height Difference of navigated vs fixed GCP points

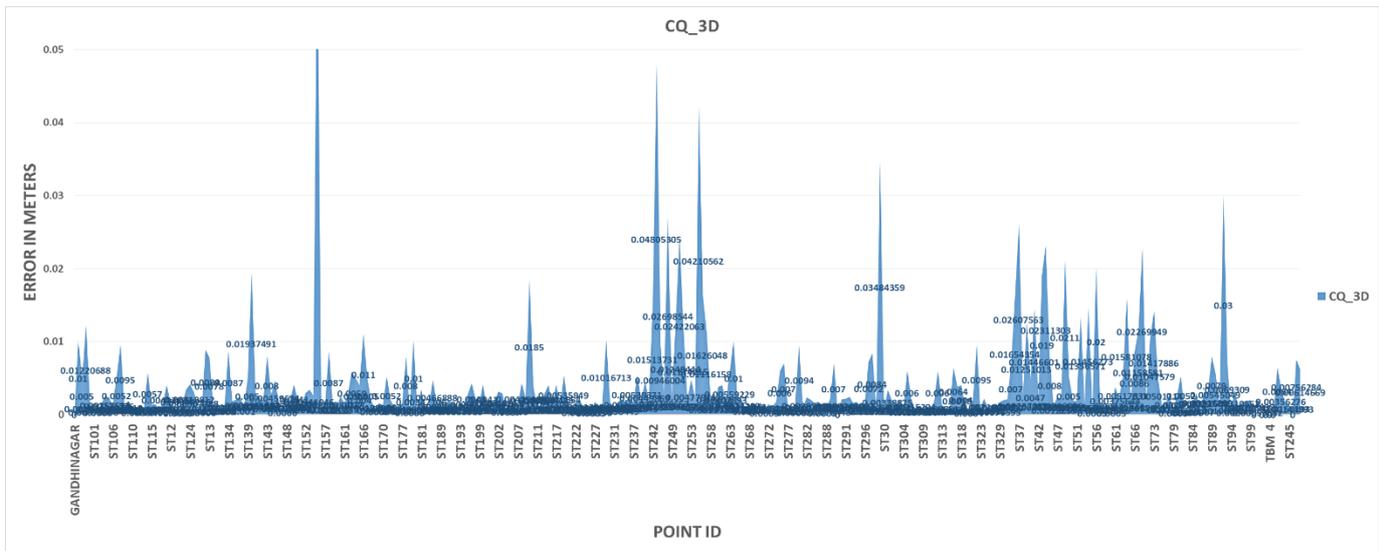

Figure-8 Error in 3D (height) values over the >300 points

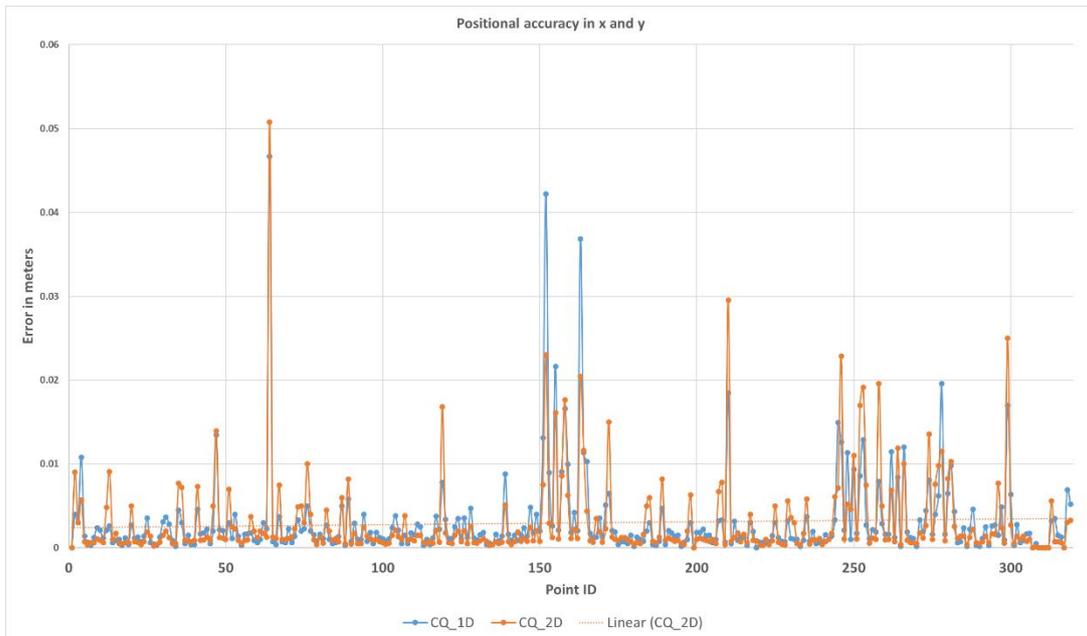

Figure-9 Positional Error in 2D (X, Y) values over the >300 points

## 5. Conclusions

A DGNSS survey over 300 points was conducted over an urban city of India in order to evaluate positional and vertical accuracies. An urban has many error sources like clock error, orbit ephemeris, ionosphere, troposphere, and multipath. But, multipath error is most prominent kind of error encountered in dense urban area. To handle all the prominent errors, proper identification, meticulous planning & surveying and post processing of the ground control point is required. In this study, we successfully performed static measurement and processing of all the points and obtained positional and vertical accuracies of millimeter to centimeter level. Precise coordinates obtained will be very handy in future applications generation of high resolution digital elevation models (DEM), validation of DEM, city modelling, planning of infrastructure in city area, change detection and many more.